\documentclass[aps,prl,twocolumn,showpacs,superscriptaddress]{revtex4}

\usepackage{graphicx}         
\usepackage{bm}               
\usepackage{amssymb}          
\usepackage{amsmath}          
\usepackage{CJK}              

    \newcommand {\fref}[1]     {Fig.~\ref{#1}}

\newcommand {\fsref}[1]    {Figs.~\ref{#1}}

\newcommand {\be}          {\begin{equation}}
\newcommand {\ee}          {\end{equation}}

\usepackage{color}

\begin{document}
\begin{CJK*}{UTF8}{bkai}

\title{Multiple Images and Flux Ratio Anomaly of Fuzzy Gravitational Lenses}

\author{James~H.~H.~Chan (詹弘旭)}
\affiliation{Institute of Physics, Laboratory of Astrophysique, \'Ecole Polytechnique F\'ed\'erale de Lausanne (EPFL), Observatoire de Sauverny, 1290 Versoix, Switzerland}

\author{Hsi-Yu~Schive (薛熙于)}
\affiliation{Department of Physics, National Taiwan University, Taipei 10617, Taiwan}
\affiliation{Institute of Astrophysics, National Taiwan University, Taipei 10617, Taiwan}
\affiliation{Center for Theoretical Physics, National Taiwan University, Taipei 10617, Taiwan}
\affiliation{Physics Division, National Center for Theoretical Sciences, Hsinchu 30013, Taiwan}

\author{Shing-Kwong~Wong (黃承光)}
\affiliation{Department of Physics, National Taiwan University, Taipei 10617, Taiwan}

\author{Tzihong~Chiueh (闕志鴻)}
\email{chiuehth@phys.ntu.edu.tw}
\affiliation{Department of Physics, National Taiwan University, Taipei 10617, Taiwan}
\affiliation{Institute of Astrophysics, National Taiwan University, Taipei 10617, Taiwan}
\affiliation{Center for Theoretical Physics, National Taiwan University, Taipei 10617, Taiwan}

\author{Tom Broadhurst}
\affiliation{Department of Theoretical Physics, University of the Basque Country, UPV/EHU, Bilbao, Spain}
\affiliation{Donostia International Physics Center (DIPC), 20018 Donostia, The Basque Country}
\affiliation{Ikerbasque, Basque Foundation of Science, Bilbao, Spain}

\date{\today}

\begin{abstract}
 Extremely light bosonic wave dark matter ($\psi$DM) is an emerging dark matter candidate contesting the conventional cold dark matter paradigm and a model subject to intense scrutiny of late.  This work for the first time reports testable salient features pertinent to gravitational lenses of $\psi$DM halos.  $\psi$DM halos are distinctly filled with large-amplitude, small-scale density fluctuations with $\delta\rho/\rho_{\rm halo}\sim 1$ in form of density granules.  This halo yields ubiquitous flux ratio anomalies of a few tens of percent, as is typically found for lensed quasars, and may also produce rare hexad and octad images, for sources located in well-defined caustic zones.  We have found new critical features appearing in the highly de-magnified lens center when the halo has sufficiently high surface density near a very compact massive core.   
\end{abstract}

\maketitle
\end{CJK*}

\noindent{\it Introduction---}
Wave dark matter ($\psi$DM) model is motivated in part by string theories engaging axion-like particles \cite{hui}, and in part by solving small-scale crises in local galaxies \cite{hu,str,boy,Nature,sper,Chen1}.  $\psi$DM predicts several novel features absent in the conventional cold dark matter (CDM) model.  The linear matter power spectrum emerged from the radiation era in the early universe has a sharp cutoff at short wavelength \cite{zhang1}, thereby suppressing dwarf galaxies \cite{hu,sch2016,mocz,leu}.  In a nonlinear axion $\psi$DM model, its linear power spectrum may even overpower that of CDM immediately before the cutoff \cite{zhang1,zhang2,mexi}, which mitigates the Lyman-$\alpha$ tension \cite{leong,lyman1,lyman2} and predicts abundant mid-sized (few $10^{10}M_\odot$) dwarfs as first-generation galaxies, for which frequent major mergers are a major mode of galaxy growth \cite{sch2017}. 

$\psi$DM halos form through aggregation of matter waves \cite{bon} with large-amplitude interference patterns (density granules) on typical scales of kpc in dwarf galaxies to 10 pc in massive galaxy clusters for the favorable axion mass $m_a\sim10^{-22}$ eV.  The central regions of all galaxies are also predicted to host prominent cores with densities much higher than the inner halo densities but comparable in size with halo granules \cite{prl}.  
These massive cores or solitons are stable, even in the presence of baryons \cite{chan,neim} and moderate tidal forces \cite{sch2019}.
These distinct halo features have been the foci of recent active studies \cite{ivan, dav,tor,zan,bar,bar1,Lee,chur,baror,mar,lev,li}. In this work, we shall report new features of strong gravitational lensing arising from halo turbulence\cite{chiueh}.

\begin{figure}
\centering
\includegraphics[scale=0.23]{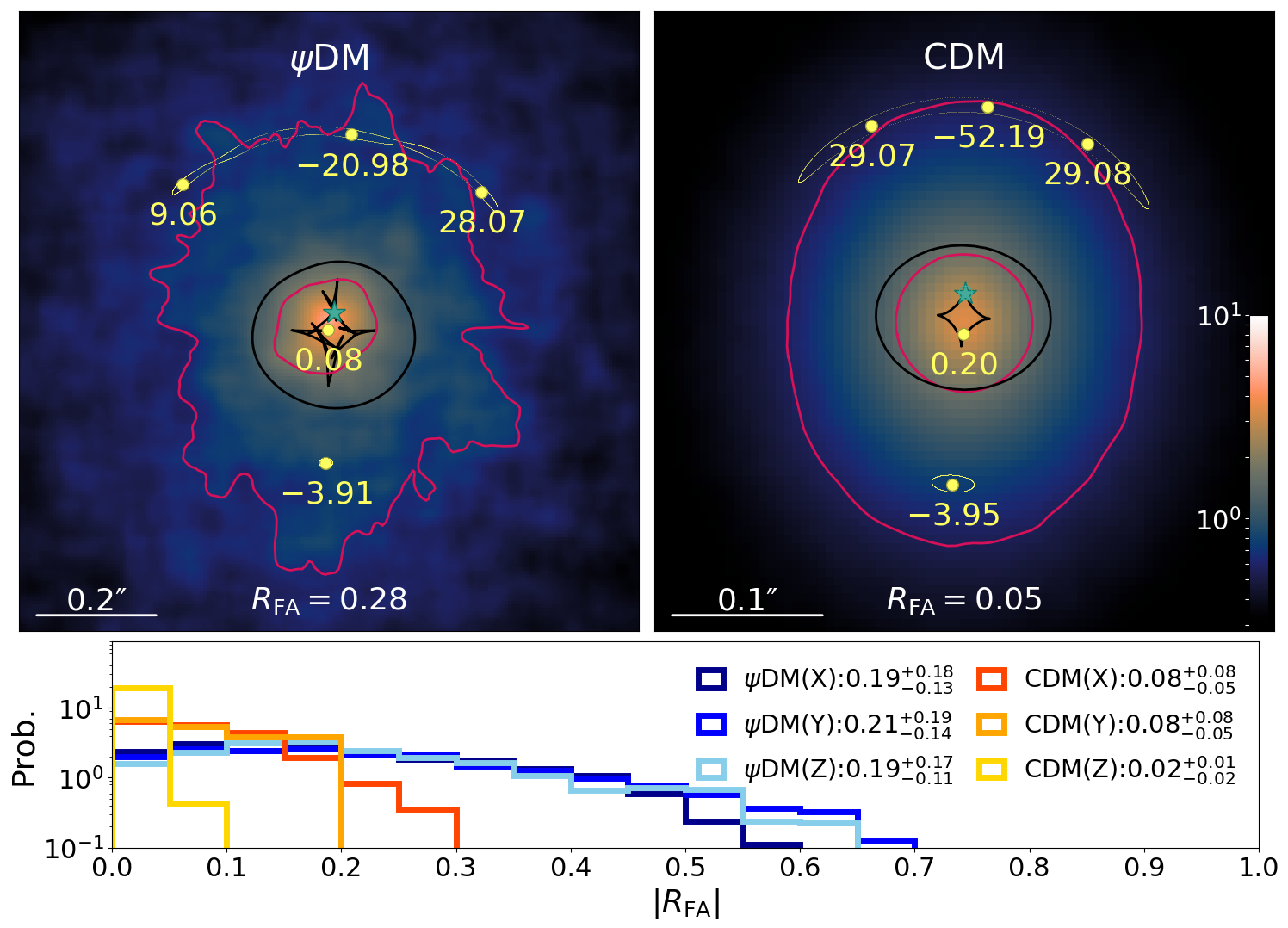}
\caption{
Surface densities normalized to the critical density (color scale) of a $z_l=0.3$ fuzzy lens (top-left) and a smooth lens (top-right) of $~10^{11.2} M_\odot$, overlaid with critical curves (red) and the corresponding caustics (black).  Note the wildly fluctuating critical curve and the spiked caustic in the fuzzy lens.  A $z_s=2$ point source (green star) is placed at the top caustic cusp and yields cusp quad images where magnifications are given.  A $40$ pc circle around the source is also mapped into tangentially stretched images.   The probability distributions of $|R_{\rm FA}|$ (bottom) show the flux ratio anomalies to be $|R_{\rm FA}|=0.19, 0.21, 0.19$ and $|R_{\rm FA}|=0.08, 0.08, 0.02$ along three orthogonal sightlines for fuzzy and CDM lenses.
}
\label{fig:flux_anomaly}
\end{figure}

\noindent{\it Quasar quad image flux anomalies---}
Strongly lensed quasar yields an odd number of multiple images including a highly de-magnified unobservable central image; therefore even numbers of lensed images are often referred.  Highly magnified images are tangentially or radially stretched on the tangential critical curve (Einstein ring) or the radial critical curve when sources are near the respective caustics.  The tangential caustic is of a diamond shape with four cusps. (See \fref{fig:flux_anomaly}.)  When a source is located near the cusp, it produces a cusp quad configuration of images consisting of three highly magnified images lying close to each other on one side of the critical curve and the fourth less magnified image on the other side.  The three bright images of cusp quads are known to obey approximate flux conservation in smooth lenses \cite{Sch, Con}.  

When subhalos are near any of the three bright images, the conservation can be broken, yielding flux ratio anomalies.  Quasar flux ratio anomalies have in past decades been employed for investigating CDM subhalos \cite{met,Kee,Dal,Mao} and recently used to constrain the warm dark matter model \cite{hsu,Gil}.  However, subhalos are rare near the small Einstein ring, and so these subhalos are suspected to be sparsely distributed outside lens galaxies intercepting the lensed images along sightlines \cite{met1,Ino,Xu}.  Consequently flux ratio anomalies can at best be moderate for CDM lenses.  By contrast $\psi$DM halos consist of granular density irregularities of finite granule mass.  Such gravitational lenses are far from smooth, more so near Einstein rings, and hence the flux ratio anomalies are expected to be a rule rather than an exception.  However to what degree is the flux conservation to be violated?  In this work we will use violation of flux conservation for cusp quads as a convenient measure for flux ratio anomalies of generic quads, as halo fluctuations are everywhere around the Einstein ring and affect all kind of bright quads to a comparable degree.

Unlike CDM simulations, the state-of-art $\psi$DM cosmological simulations are limited only to simulation boxes of a few Mpc, due to very high spatial and time resolutions required, and therefore massive lenses are difficult to come by.  One attempt has been to construct 3D massive halos by solving for all halo eigenmodes consistent with the self-gravity \cite{lin}.  This is a nonlinear self-consistent solution using iteration for the construction of the entire halo including the halo fluctuations.

However even a Milky Way sized halo requires tens of thousand pixels to resolve density granules in each dimension for halo construction, thus a formidable task.  Given this limitation, we select the most massive halo $\sim10^{11.2}M_\odot$ from $m_a= 0.8\times 10^{-22}$ eV concordance cosmology $\psi$DM simulations \cite{prl} as the lens.  We note a small halo has a large granule size $l_\sigma\propto M_h^{-1/3}$, and small virial radius $R_{\rm vir}$, and a small granule number $2R_{\rm vir}/l_\sigma$ in projection yields large lens surface density fluctuations that distort the critical curve.  By contrast too small a halo cannot exceed the critical density to produce an Einstein ring.  It turns out that a lens of $\sim 10^{11.2}M_\odot$ has a sufficient central surface mass density to generate a small Einstein ring, and is maximally fuzzy, with the largest excursion of the critical curve due the combination of larger granule size and smaller projected granule number.

This halo has a central soliton of core radius $r_c\sim 300$ pc and mass $\sim 10^{8.5}M_\odot$ \cite{Nature, prl}, and granules of mass $10^{6}M_\odot$ in the inner halo.  The halo is elliptical with $2:1.5:1$ axial ratio and we choose projection along the major axis.  Baryons are also considered, following an elliptical de Vaucouleurs profile to mimic a galaxy of $1.5$ kpc minor-axis half-light radius within which the baryon fraction is $\sim25\%$ and a  galaxy total baryon fraction of $\sim5\%$ for the galaxy as a whole.
Although the addition of a smooth stellar component smooths out the total density, making the granule effect less prominent, the dark matter density is still fully modulated due to self interference when stars and $\psi$DM are evolved self-consistently \cite{chan,neim}.  We have also conducted a corresponding CDM simulation and identified the same halo, which is $10\%$ less massive and somewhat less elliptical, for constructing a control CDM lens to compare. 
Given the surface density, the critical curves are first identified on the lens plane at $z_l=0.3$ and the caustics on the source plane at $z_s=2$ are then identified by back-ray-tracing.  We select a finite-size source to construct multiple images by forward-ray-tracing.  Shown in \fref{fig:flux_anomaly} are surface densities overlaid with sources and quad images for both fuzzy and smooth lenses, where the Einstein radii are $\theta_{\rm Ein}\sim 0.3$ arcsec ($\sim 1.25$ kpc) and $0.13$ arcsec, respectively.  The magnification $\mu$ is proportional to the local area of the distorted image, and $\mu$ is negative when the image reverses its parity occurring inside the critical curve.  Compared with the smooth lens, the fuzzy lens has a wildly fluctuating tangential critical curve and the diamond caustic exhibits unusual spikes.  We note that fluctuating critical curve makes its distances from images randomly vary, therefore resulting in flux anomaly. 

The degree of flux ratio anomaly of the cusp images is defined as $R_{\rm FA}=\sum_j\mu_j/\sum_j|\mu_j|$, and $j=1,2,3$ for the bright triplet.  Accounting for all sources located near four diamond cusps with $\sim 10^3$ sources, we find the mean $\langle|R_{\rm FA}|\rangle=0.19^{+0.18}_{-0.13}, 0.21^{+0.19}_{-0.14}, 0.19^{+0.17}_{-0.11}$ (median and ($-$)16th($+$)84th percentiles) for the fuzzy lens and $\langle|R_{\rm FA}|\rangle=0.08^{+0.08}_{-0.05}, 0.08^{+0.08}_{-0.05}, 0.02^{+0.01}_{-0.02}$ for the smooth lens along three orthogonal axes.  Highlighted in \fref{fig:flux_anomaly} are typical images of fuzzy and smooth lenses with $R_{\rm FA}=0.28$ and $R_{\rm FA}=0.05$, respectively.  This result leads us to conclude that flux ratio anomalies can indeed be sizable for $10^{11.2}M_\odot$ fuzzy lenses.

\noindent{\it Hexad and octad images---}
\begin{figure}
\centering
\includegraphics[scale=0.3]{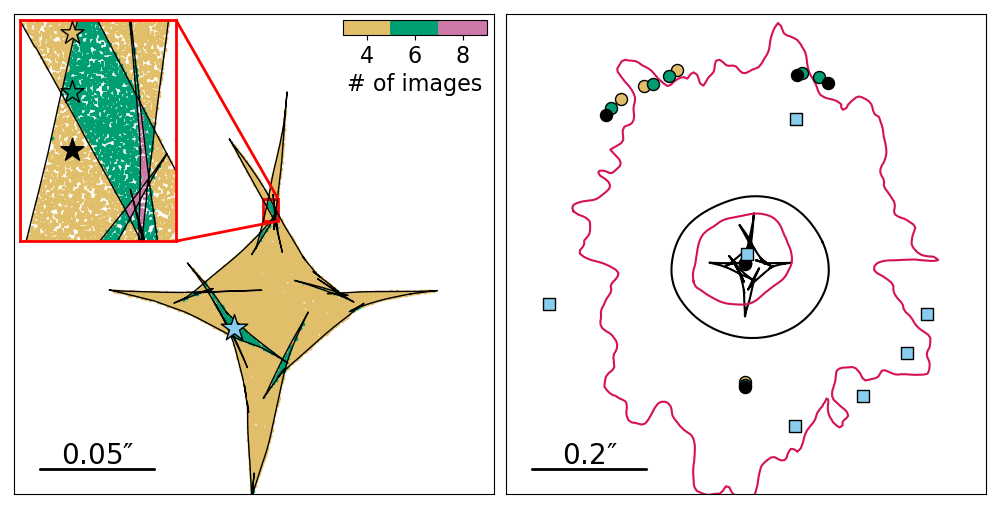}
\caption{
{\it Left:} A zoom-in map of the spiked caustic with a further blow-up view of a diamond cusp to show zones (brown, green, pink) of different image numbers, where four point sources (stars) are placed in different zones.  Note that the caustic spikes emerge from the joints of cusps and folds.  {\it Right:} Cusp and fold images produced by these point sources inside and outside the blow-up region are also shown.   
}
\label{fig:multi_imgs}
\end{figure}
When a point source is located inside some well-defined zones of the diamond caustic spikes, it may yield magnified 6 images or 8 images.  These new images appear or disappear when a source crosses caustics.  The right panel of \fref{fig:multi_imgs} highlights cusp and fold hexad images (green circles and light blue circles) as well as two cusp quads (brown circles and black circles).  The cusp hexad has $5$ bright images on the near side of critical curve with magnifications $\mu=(19.0,-27.3,39.5,-33.7,41.6)$ from top left to top right.  The left panel of \fref{fig:multi_imgs} is the zoom-in of a caustic cusp, where different color zones represent different image numbers.  A cusp hexad can transition to two different quads (right panel) when the source crosses different zone boundaries (left panel).  The number of images changes in pairs, which either appear or disappear on crossing a caustic boundary. The brightnesses of these image pairs are similar and become very bright when close to the caustic.  A fold hexad can transition to a fold quad in a similar manner.

The probabilities of these new images are proportional to respective caustic zonal areas.  We have computed the probability ratios of quads, hexads, and octads to be $1: 7.5\%: 0.1\%$, assuming point sources.  However for very faint sources, detectable images are biased to be located along the caustic boundary to be highly magnified.  Hence we estimate boundary lengths as the measure of probability for highly magnified images.  Here one boundary is shared by two types of images.  Highly magnified quad, hexad, and octad probability ratios now roughly become $1:28\%:0.8\%$ as a result of drastic reduction of highly magnified quads. 

\noindent{\it Extension to massive fuzzy lenses---}
\begin{figure}
\centering
\includegraphics[scale=0.24]{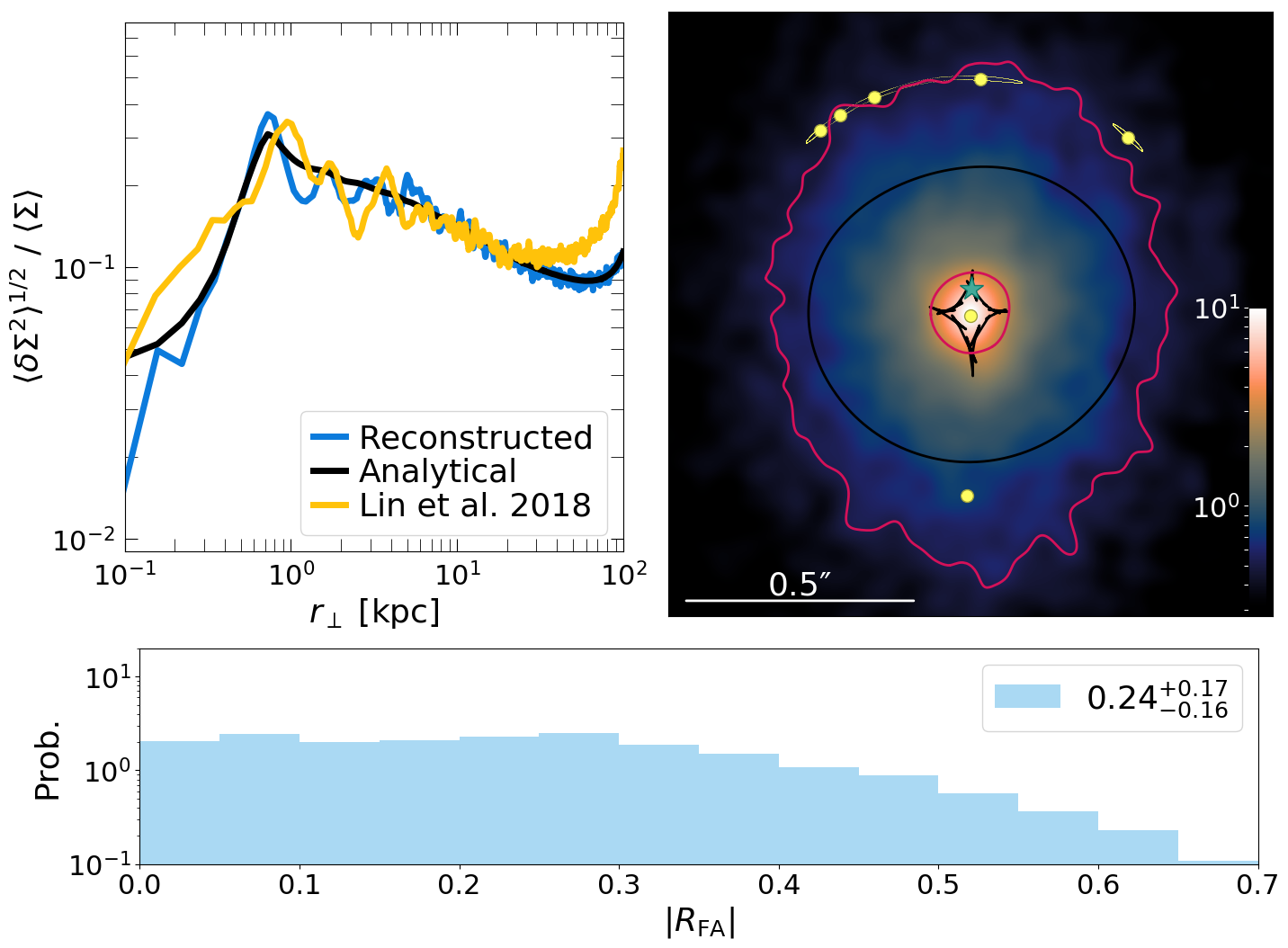}
\caption{
(Top-left) Predicted (black) and reconstruction (blue) profiles of the surface fluctuation, $\langle\delta\Sigma^2\rangle^{1/2}/\langle\Sigma\rangle$ (c.f., the formula given in the text) compared with that of a control halo of $10^{10.8}M_\odot$ (yellow) to show good agreement in the inner halo.  (Top-right) A reconstructed $10^{11.8}M_\odot$ elliptical lens similar to \fref{fig:flux_anomaly}, which yields a cusp hexad.  The critical curve fluctuation is clearly smaller than, but the $|R_{\rm FA}|$ probability distribution (bottom) is comparable to, those of the less massive lens of \fref{fig:flux_anomaly}.
}
\label{fig:reconstruced}
\end{figure}
Given the difficulty in obtaining massive halos via simulations, we describe how to construct a fuzzy lens of a more massive halo or a fuzzy lens of larger $m_a$.  The fluctuating halo wave function $\psi(x)$ can be Gaussian random, and since $\rho_\psi({\bf r})=m_a|\psi({\bf r})|^2$, it yields a random density obeying an exponential density distribution, $P(\rho_\psi)=(1/\langle\rho_\psi({\bf r})\rangle)\exp[-\rho_\psi({\bf r})/\langle\rho_\psi({\bf r})\rangle]$ with a mean $\langle\rho_\psi({\bf r})\rangle$ and variance $\langle\delta\rho_\psi^2\rangle=\langle\rho_\psi({\bf r})\rangle^2$.  We further assume the Maxwell-Boltzmann energy distribution for the random phased classical field \cite{lev}, $|\psi({\bf k})|^2 \propto \exp[-(1/2)(\hbar k/m_a\sigma_v)^2]$, which is consistent with the Gaussian statistic of $\psi({\bf r})$, the inverse Fourier transform of $\psi({\bf k})$.  Here $\sigma_v$ is the $1$D halo velocity dispersion.  From the energy distribution and Gaussian random $\psi$, it follows $|\rho_{\psi}({\bf k})|^2\propto\exp[-(1/4)(\hbar k/m_a\sigma_v)^2]$ and the density correlation $\langle\delta\rho_\psi({\bf x})\delta\rho_\psi({\bf x}+{\bf r})\rangle_{\bf x}=\langle\delta\rho^2_\psi({\bf x})\rangle \exp[-(r/l_\sigma)^2]$ where the correlation length $l_\sigma=\hbar/m_a\sigma_v$.  Not surprisingly, the core radius $r_c$ of the central soliton \cite{Nature, prl} is almost identical to $l_\sigma$.
    
On projecting the 3D halo into a lens, the central limit theorem demands the surface density fluctuations, $\delta\Sigma({\bf r}_\perp)=\int \delta\rho_\psi({\mathbf{r}})dz$, to be Gaussian random with the variance $\langle\delta\Sigma^2({\bf r}_\perp)\rangle=\int\int \langle\delta\rho_\psi({\bf r})\delta\rho_\psi({\bf r}+{\bf z}')\rangle dz dz'=\int dz\langle\delta\rho_\psi^2({\bf r})\rangle\int dz'\exp[-(z'/l_\sigma)^2]=\int dz\langle\rho_\psi({\bf r})\rangle^2(\sqrt{\pi}l_\sigma) =\sqrt{\pi}\langle l_\sigma\rangle\int dz \langle\rho_\psi({\bf r})\rangle^2$, where ${\bf r}_\perp\perp \hat z$.  Note that the squared density weighted correlation length $\langle l_\sigma\rangle=l_\sigma$ if $\sigma_v({\bf r})$ is constant over the entire halo.  But this is only valid when $r < 2 r_s$ where $r_s$ is the scale radius of the halo beyond which the density drops rapidly but $\sigma_v$ declines slowly, as shown the gradual increase of granule size \cite{lin}.  Nevertheless, $\langle l_\sigma\rangle$ is strongly biased against contributions outside $r_s$ since $\langle\rho_\psi\rangle^2$ decreases outward rapidly and moreover strong lensing mostly occurs inside $r_s$.  Hence we take $\langle l_\sigma\rangle=l_\sigma$.  The left panel of \fref{fig:reconstruced} displays the theoretically predicted and reconstructed surface fluctuations $\langle\delta\Sigma^2\rangle^{1/2}/\langle\Sigma\rangle$ for constant $\langle l_\sigma\rangle$ following $\langle\rho_\psi\rangle$ of a $10^{10.8}M_\odot$ self-consistent halo with $r_s\sim 8$ kpc \cite{lin}.  This halo, whose actual surface density is to compare, is spherical, consists of random-phased halo eigenfunctions and has been tested to be stable against the simulation.  The comparison shows good agreement for $r_\perp < 2r_s$.  Beyond $2r_s$, the 3D halo has greater surface fluctuations since $\sigma_v$ becomes smaller and $\langle l_\sigma\rangle$ larger.  The central depletion is due to the presence of a prominent coherent soliton.

Having confirmed the method, we construct a fuzzy lens for a $8\times 10^{11}M_\odot$ elliptical galaxy with the axial ratio $1.3:1:1$ inside the $5$ kpc half-light radius.  The NFW halo profile, which $\psi$DM halos largely follow \cite{Nature, prl}, is adopted for lens construction where $r_s=13$ kpc and the virial radius $200$ kpc.  The lens is made elliptical via uniform stretch along the major axis. The baryon fraction is $50\%$ within the half-light radius and $10\%$ over the entire galaxy.  The right panel of \fref{fig:reconstruced} shows the dark matter surface density of this medium-mass lens with $m_a=0.8\times 10^{-22}$ eV, overlaid with critical curves and caustics.  Again $z_l=0.3$ and $z_s=2$.  The wiggly critical curve is similar to that of the $10^{11.2}M_\odot$ lens, but fluctuating with smaller amplitudes and on a smaller scale ($200$ pc); the caustic cusps also possess similar, but somewhat weaker, spikes.  The $|R_{\rm FA}|$ probability distribution is however found to be comparable to the less massive lens, with the mean $\langle|R_{\rm FA}|\rangle =0.24^{+0.17}_{-0.16}$.  Also shown in \fref{fig:reconstruced} is the hexad image, the salient feature found in less massive lenses which also appears in the $8\times10^{11}M_\odot$ lens.  But hexad images have a smaller probability $3.5\%$ relative to quad images than the less massive lens as a result of reduced surface fluctuations; the probability of octad images is even smaller to a level beyond our numerical resolution.

For even more massive lenses, the wavelength and amplitude of fluctuations in the critical curve are both reduced as $M_h^{-1/3}$, suggesting that the degree of curvature of the critical curve will be of a similar scale over the full range of lensing mass.  While hexad images are to diminish in a more massive lens, the flux anomaly is expected to decrease at a much slower rate since these cusp images are located in close proximity to the critical curve and sensitive to the fluctuations.  This tendency of flux anomaly for very small-size granular fluctuations can be illustrated in the ensuing investigation where the axion mass is increased by $10$ times.
\begin{figure}
\centering
\includegraphics[scale=0.23]{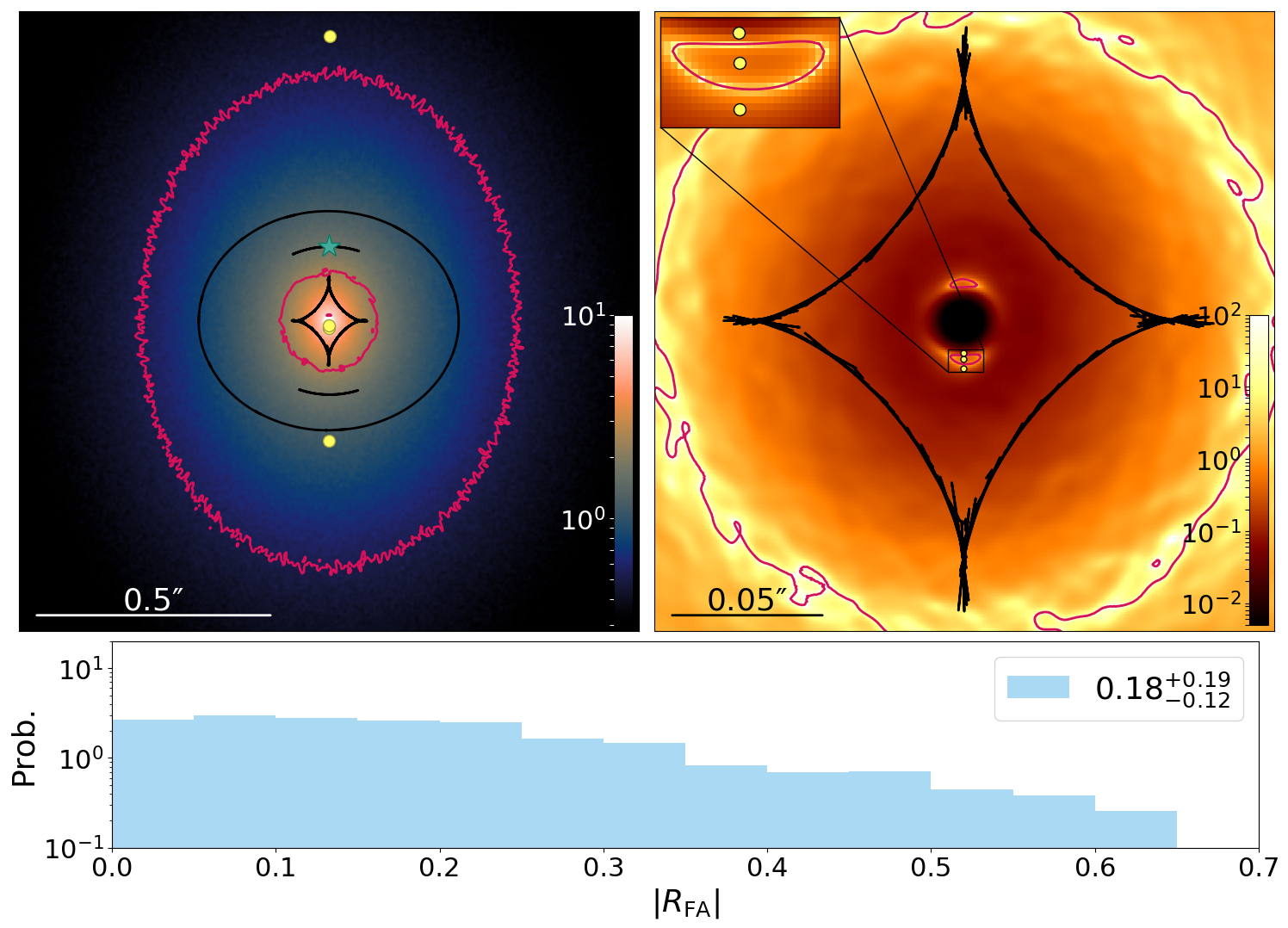}
\caption{
Fuzzy lens similar to \fref{fig:reconstruced} but with the axion mass, $m_a=8\times 10^{-22}$ eV.  The wavelength and amplitudes of fluctuations on the critical curve are much smaller than those in \fref{fig:reconstruced} and spikes in the caustic now become bristle-like.  However the flux anomaly distribution (bottom) still resembles those of other lenses shown in \fsref{fig:flux_anomaly} and \ref{fig:reconstruced}.  In addition, new critical features emerge, where three images of a quad are packed around one central critical loop (top-right) when the source is almost on a thin caustic (top-left).   The color scale in the top-right panel represents the magnification $|\mu|$ showing large $\mu$ gradient near critical loops.
}
\label{fig:innerloop}
\end{figure}
Employing the lens reconstruction method for the same $8\times 10^{11} M_\odot$ elliptical galaxy but with $m_a=8\times 10^{-22}$ eV, we show results in \fref{fig:innerloop}. Clearly seen are the wiggle wavelength and amplitude of the tangential critical curve reduced by $m_a^{-1}$ and $m_a^{-1/2}$, respectively. The critical curve appears highly distorted with large local curvatures.  Moreover, caustic spikes are now bristle-like spreading over the entire diamond for which zones of higher image numbers almost disappear.  Despite that, the flux anomaly remains comparable to other lenses discussed previously with $\langle|R_{\rm FA}|\rangle = 0.18^{+0.19}_{-0.12}$.  

Besides, new features emerge in \fref{fig:innerloop}.  They are two additional small critical loops located at the edge of the prominent core and two corresponding very thin caustic loops.  When the source is almost on the new caustic, the image is an unconventional quad where three images are centrally packed about one small critical loop and other two moderately magnified images are located inside and outside the tangential critical curve.  With tiny source displacements the triplet can either be all de-magnified resembling that of a double image, or have two almost overlap bright images (radial arc) and one de-magnified image.  It can be so sensitive because the magnification gradient is enormous there as shown in the left panel of \fref{fig:innerloop}.

The new critical features arise in the presence of a very compact massive core (soliton) along with high central surface density.  For a circularly symmetric lens, they correspond to new zeros of the inverse determinant of the deformation matrix \cite{schneider}, proportional to $[\Bar{\kappa}-\kappa]-(\kappa-\kappa_{\rm sol})+1$, where $\kappa$ is the surface density $\Sigma$ divided by the critical density $\Sigma_{\rm cr}$, $\Bar{\kappa}$ the enclosed average $\kappa$ within $r$ and $\kappa_{\rm sol}=(M_{\rm sol}/\pi r^2)\Sigma_{\rm cr}^{-1}$.  At small $r$ where $\kappa \gg 1$, the square bracket is small and the new zeros are located near $\pi r^2\Sigma(r)= M_{\rm sol}$ as is found here. By contrast, the ordinary radial critical curve corresponds to the zero of $\kappa=O(1)$, which happens to be the only zero for the lenses discussed previously due to their wide cores to yield insufficiently large $\kappa$ throughout these lenses.  For $m_a=0.8\times 10^{-22}$ eV, this new zero, contributed by a very compact core, can only occur in lenses $M_h>10^{14}M_\odot$ (inverse soliton radius and soliton mass $\propto M_h^{1/3} $\cite{prl}). Nevertheless a less massive halo that hosts a super-massive black hole inside the soliton may also generate this novel feature.

\noindent{\it Concluding remarks---}
We have shown salient strong lensing features in $10^{11.2} M_{\odot}$ and Milky Way sized ($10^{12} M_\odot$) sized $\psi$DM lenses.  The former are predicted to be about several times more abundant than the latter according to the mass function of galaxies \cite{sheth, and  therefore can have comparable optical depths for lensing background sources.}  Moreover once intercepting, they are more probable to yield hexad images, despite their small Einstein radii that require adaptive optics observations \cite{chen1,Chen}.  Nevertheless surveillance-type low-resolution observations may prove to be useful to help detecting the hexad images found in this work.  Rapid transients, such as fast radio bursts \cite{Tho}, $\gamma$-ray bursts \cite{Berger}, and stellar gravitational waves \cite{abb}, are originated from bright distant point sources.  The multiple lensed images of these sources can result in repetitions of the transient signal due to time delays among images \cite{suyu}.  The inter-burst intervals of repetition can range from minutes to hours for a $10^{11.2}M_\odot$ lens, and from an hour to a day for a $10^{12} M_\odot$ lens when a source is located at $z=1$.  

Flux ratio anomalies have been detected for most quasar quads over a wide spectrum.  Optical and x-ray \cite{Irw, Poo1,Poo2} flux ratio anomalies have largely been attributed to dust extinction and microlensing on different image pathways.  However in infrared and submillimeter, the dust extinction and microlensing are lesser a problem \cite{Min, Jack}, but even in these wave bands, quasar quads, such as MG J0414+0534, B1422+231, H1413+117, and B2045+265 in radio, submillimeter, mid-infrared \cite{Sta, Mac2, Mac1, Chi} and/or infrared \cite{Mck} as well as other quads in narrow lines \cite{nie} persistently show substantial flux ratio anomalies.  Aside from the conventional interpretation of intercepting subhalos, the fuzzy lens of $\psi$DM can offer a compelling explanation if not more so, given the $10-30\%$ flux ratio anomaly to be a norm and not an oddity.

\section*{Acknowledgments}
J.~H.~H.~C.~acknowledges support from the Swiss National Science Foundation (SNSF).  H.~S.~acknowledges the funding support from the Jade Mountain Young Scholar Award No. NTU-108V0201, sponsored by the Ministry of Education, Taiwan.  This research is partially supported by the Ministry of Science and Technology (MOST) of Taiwan under the grant No.  MOST 107-2119- M-002-036-MY3 and MOST 108-2112-M-002-023-MY3, and the NTU Core Consortium project under the grant No. NTU-CC-108L893401 and NTU-CC-108L893402.


\end{document}